\documentclass[a4paper,11pt]{article}
\pdfoutput=1 

\usepackage{jinstpub} 

\title{RFSoC-based front-end electronics for pulse detection}

\author[d]{S.~N.~Axani}
\author[a,1]{S.~Futagi\note{Present address: National Institute of Advanced Industrial Science and Technology, Tsukuba, Ibaraki 305-8560, Japan}}
\author[d]{M.~Garcia}
\author[c]{C.~Grant}
\author[a,2]{K.~Hosokawa\note{Present address: Kamioka Observatory, Institute for Cosmic-Ray Research, The University of Tokyo, Hida, Gifu 506-1205, Japan}}
\author[a]{S.~Ieki}
\author[a]{K.~Inoue}
\author[a]{K.~Ishidoshiro}
\author[a]{N.~Kawada}
\author[a]{Y.~Matsumoto}
\author[a]{T.~Nakahata}
\author[a]{K.~Nakamura}
\author[a]{R.~Shouji}
\author[c]{H.~Song}
\author[b]{L.~A.~Winslow}

\affiliation[a]{Research Center for Neutrino Science, Tohoku University, Sendai 980-8578, Japan}
\affiliation[b]{Massachusetts Institute of Technology, Cambridge, Massachusetts 02139, USA}
\affiliation[c]{Boston University, Boston, Massachusetts 02215, USA}
\affiliation[d]{University of Delaware, Newark, Delaware 19711, USA}

\emailAdd{koji@awa.tohoku.ac.jp}

\abstract{Radiation measurement relies on pulse detection, which can be performed using various configurations of high-speed analog-to-digital converters (ADCs) and field-programmable gate arrays (FPGAs). For optimal power consumption, design simplicity, system flexibility, and the availability of DSP slices, we consider the Radio Frequency System-on-Chip (RFSoC) to be a more suitable option than traditional setups. To this end, we have developed custom RFSoC-based electronics and verified its feasibility. The ADCs on RFSoC exhibit a flat frequency response of 1-125 MHz. The root-mean-square~(RMS) noise level is 2.1\,ADC without any digital signal processing. The digital signal processing improves the RMS noise level to 0.8\,ADC (input equivalent 40\,$\mu$V$_\textrm{rms}$). Baseline correction via digital signal processing can effectively prevent photomultiplier overshoot after a large pulse. Crosstalk between all channels is less than -55\,dB. The measured data transfer speed can support up to 32 kHz trigger rates~(corresponding to 750~Mbps). 
Overall, our RFSoC-based electronics are highly suitable for pulse detection, and after some modifications, they will be employed in the Kamioka Liquid Scintillator Anti-Neutrino Detector (KamLAND).

}

\keywords{Front-end electronics for detector readout, Digital signal processing (DSP)}

\begin{document}
\maketitle
\flushbottom

\section{Introduction}
\label{sec:intro}
Pulse detection serves as a fundamental role in particle and nuclear physics~\cite{Leo:1993xta}, and radiation measurements~\cite{Knoll:2010xta}. 
In the pursuit of extracting maximal information from pulses, the acquisition of pulse waveforms stands as a crucial objective. 
To address this goal, systems equipped with high-speed analog-to-digital converters~(ADCs) and field-programmable gate arrays~(FPGAs) are commonly employed. 
In most cases, ADCs digitize the analog waveforms with a sampling speed of 50~MS/s to 1~GS/s and a resolution of 8 to 12-bits.  
The FPGA is used mainly to control and read out the ADCs and transmit the digitized waveforms to back-end computers.
To improve sampling speed, application-specific integrated circuits~(ASICs) such as Domino Ring Sampling (DRS4) chips~\cite{Bitossi:2016waj} are often used. 
Nevertheless, the DRS4 is not ideal for dead-time free measurements.  
In astroparticle physics, dead-time free measurements assume significance, particularly in experiments targeting rare events like neutrinos, dark matter, and neutrinoless double beta decay. This is due to certain background discrimination strategies necessitating the capture of pulses very close in time. 

Consequently, many rare event search experiments devise custom electronics~\cite{Nishino:2007ccp,Bellato:2020lio, Abe:2013iha,AXEL:2020nwp}. 
Similarly, we have developed and implemented dead-time free front-end electronics within the Kamioka Liquid Scintillator Anti-Neutrino Detector (KamLAND). 
KamLAND is an 1~kiloton liquid scintillator (LS) detector located in the Kamioka mine. 
It was originally designed to detect $\bar{\nu}_e$ from Japan’s nuclear reactors~\cite{KamLAND:2013rgu} and geoneutrinos~\cite{KamLAND:2022vbm}. 
Currently, KamLAND is searching for neutrinoless double beta decay~(0$\nu \beta \beta$) of $^{136}$Xe in the KamLAND-Zen experiment. 
The dead-time free front-end electronics played a pivotal role in enabling world-leading search for $0 \nu \beta \beta$~\cite{KamLAND-Zen:2016pfg,KamLAND-Zen:2022tow}. 
Nonetheless, the front-end electronics exhibits limitations concerning buffer capacity, readout speed, and digital processing resources. 
These limitations hinder the detection efficiency of neutrons originating from cosmic-ray spallation, one of the most significant backgrounds in KamLAND-Zen.
Triple coincidences among cosmic-ray muons, neutron events, and spallation product decay can identify the spallation backgrounds. 
However, the difficulty of the triple coincidence identification is the low neutron detection efficiency. 
Cosmic-ray muons induce large pulses that saturate the Photomultiplier Tubes~(PMTs), subsequently burying the neutron-related signals within PMT overshoot and afterpulses. 
To enhance neutron detection efficiency, it is desirable to correct the waveform overshoot and capture all pulses generated, including afterpulses, with substantial buffer capacity and \textcolor{black}{high throughput} readout. 
Accompanied by state-of-the-art software processing on back-end computers, this approach can provide better neutron detection efficiency. 
Additionally, the power consumption posses additional constraints in environments with restricted power and cooling resources. 

To solve these problems, we propose the integration of Radio Frequency System-on-Chip~(RFSoC) technology. 
The RFSoC, a cutting-edge FPGA family developed by AMD Xilinx, amalgamates programmable logic~(PL) with a large number of digital signal processors, a quad-core processing system~(PS) based on Arm Cortex-A53 and a dual-core Arm Cortex-R5F processors, and radio-frequency ADCs~(RF-ADCs), and radio-frequency digital-to-analog converters~(RF-DACs). 
The application of RFSoC technology holds promise for achieving dead-time free pulse detection, particularly within the KamLAND-Zen experiment. It offers notable advantages over conventional systems, including easy implementation of high speed data transfer with 1~GbE or 10~GbE, easy handling to external and/or internal memory, low power consumption due to the tight integration between the ADCs and PL, and the large number of logic cells and digital signal processors. 
RFSoCs have begun to make inroads into physics experiments, aiding in controlling and extracting data from superconducting qubits~\citep{Stefanazzi:2021otz,Park:2021kdd}. 
Moreover, the application of RFSoC technology in radio astronomy receivers has been recently explored~\cite{Liu:2020ab}. 
Despite these advances, the application of RFSoC for pulse detection remains uncharted territory. 

In this paper, we conduct feasibility studies on a custom built RFSoC-based front-end system aimed at achieving dead-time free pulse detection.

\section{Prototype design}
\begin{figure}
\centering
\includegraphics[width=100mm]{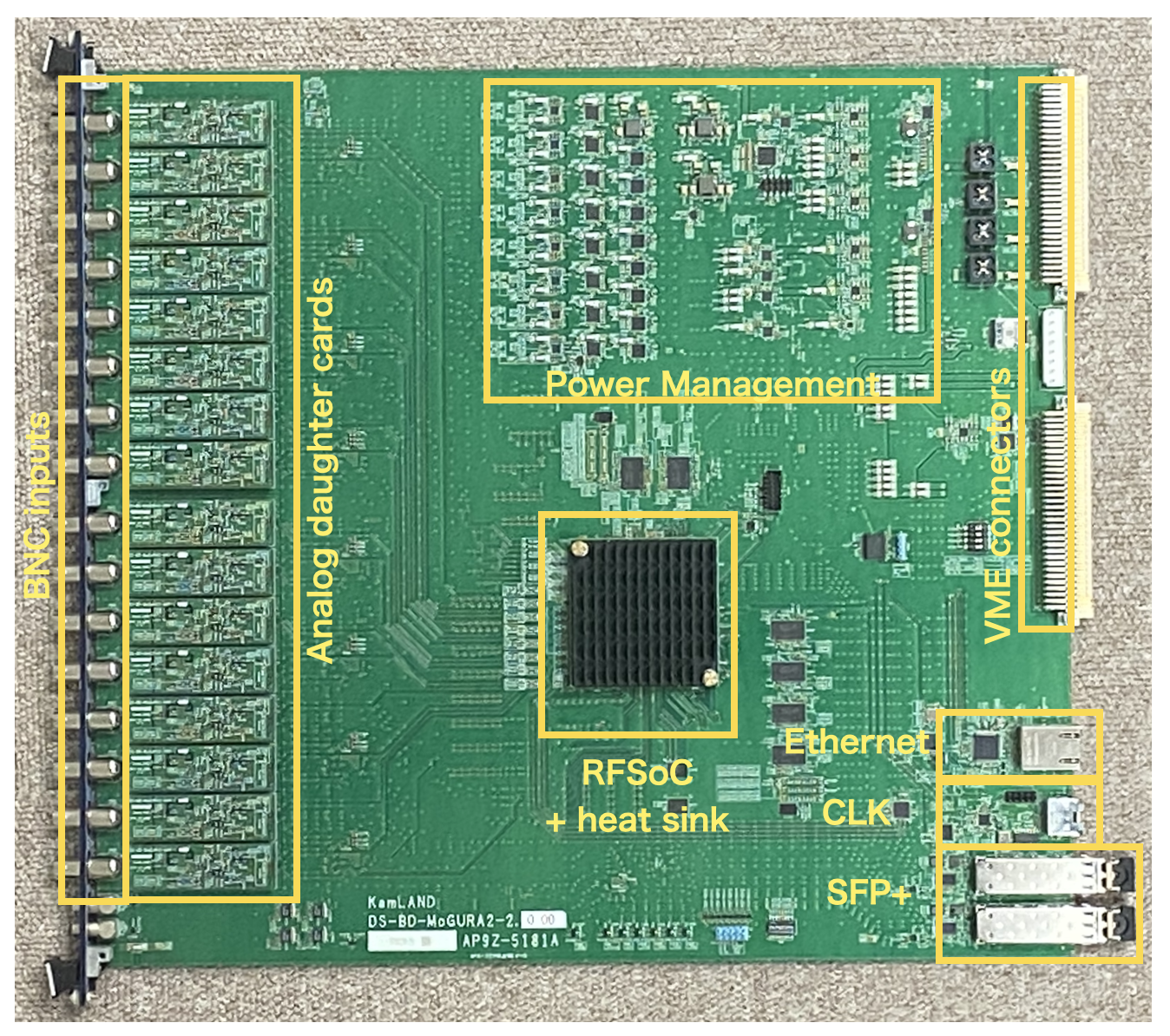}
\caption{The front-end electronics for our feasibility studies. }
\label{fig:photo}
\end{figure}

The custom electronics (shown in Fig.~\ref{fig:photo}) consists of a Main Board and 16 analog daughter cards.
The firmware and software for this board was developed with Vivado 2019.2, Vitis 2019.2, and PetaLinux2019.2. 
\textcolor{black}{This section describes the requirements and design of our RFSoC-based front-end electronics.}

\subsection{Requirements for KamLAND2}
\textcolor{black}{
The custom electronics have been designed to match the requirements for the KamLAND upgrade, referred to as KamLAND2.
}

\textcolor{black}{
The current KamLAND front-end electronics adheres to the VME9U standard~(9U $\times$ 400\,mm). The new electronics must conform to the VME9U board specifications to facilitate the reuse of the power supply units. KamLAND is in an underground laboratory, and its cooling system relies on mine water. Due to limited cooling water supply capacity, a significant reduction in power consumption per channel is strongly desired. KamLAND2 incorporates approximately 2,000 20-inch PMTs. To minimize power consumption, an augmentation of channel density is a logical consideration. The current front-end electronics supports 12 analog inputs; however, this should be increased to 16 to effectively curb power consumption.}

\textcolor{black}{
We want to analyze PMT waveforms ranging from single photoelectron~(PE) events with a typical amplitude of $\sim$2\,mV to large cosmic-ray muon events with a maximum expected amplitude of $\sim$8\,V. The analog-to-digital conversion rate must exceed 1\,GS/s for vertex reconstruction and particle identification with SPE events, but the sampling speed requirement for muon events is less stringent. 
To achieve a high signal-to-noise ratio for single PE events and establish a 1/4\,PE discrimination threshold, the noise level should be below 0.1\,mV. A single ADC cannot have the dynamic range from single PE events to large muon events with this noise level. Thus, each PMT analog input is divided into a high-gain ($h$-gain) channel for signal-to-a-few tens PE events, and low-gain ($l$-gain) channel for muon events. For consistency between single PE events to muon events, the operating ranges of $h$-gain and $l$-gain must partially overlap. To prevent charge loss, the analog and digital electronics must maintain a flat frequency response up to $\sim$60\,MHz.}

\textcolor{black}{The primary objective of the new front-end electronics is to enhance neutron detection efficiency. 
The implementation of digital signal processing to correct PMT overshoot is a key aspect of this project. 
Addressing the issue of PMT afterpulses is also crucial, since afterpulses can mask neutron events. 
As such, we aim to record all candidate neutron events, including afterpulses, and process them offline.
Assuming a single event duration of 80\,ns, translating to 184\,Bytes, including headers and footers, we anticipate an afterpulse rate of 1\,MHz lasting 1\,ms following a muon event. Consequently, the data rate related to afterpulses is about 7\,Mbps with a muon rate of 0.3\,Hz. The data rate associated with afterpulses will be easily processed with the 100\,MbE readout and the buffer which is larger than 1\,GB. 
The plan for KamLAND2 involves the use of software triggers. For the software trigger, all waveforms, including dark pulses and LS events, must be transmitted to back-end computers. With a conservative dark rate of 25\,kHz, per PMT, the minimum required data rate to record all dark hits is 588.8\,Mbps. A data transfer speed of 600\,Mbps suffices for the enhancement of the neutron detection efficiency and the software trigger. Thus, the electronics must have the 1\,GbE readout and the large buffer. }

\textcolor{black}{We have considered classical 1\,GS/s ADCs and FPGA configuration and the utilization of an 8-channel RFSoC. The former would inevitably lead to increased power consumption, and there are challenges in coordinating timing between multiple ADCs and FPGA with a sampling rate of 1\,GS/s. The latter option would either increase the number of boards or the number of RFSoCs per board, rendering it economically inefficient for handling a total of 2,000 PMTs. Thus, we use a 16-channel RFSoC for our front-end electronics.}

\subsection{Main Board}
The main board is designed in accordance with the VME9U standard. 
The main board accommodates a single RFSoC (XCZU29DR-1FFVF1760E), hosting 16\,RF-ADCs with 12-bit resolution and a maximum sampling rate of 2.058\,GS/s, and 16~RF-DACs with 14-bit resolution and a maximum sampling rate of 6.554\,GS/s. This RFSoC also integrates 930,300 logic cells, 4,272 digital signal processing elements (DSP slices), a quad-core Arm Cortex-A53, and a dual-core Arm Cortex-R5. The PS and PL are each integrated with 4\,GB Double-Data-Rate~(DDR) Synchronous Dynamic Random Access Memory~(SDRAM).  


The front panel accommodates 16 Bayonet-Neel-Concelman~(BNC) connectors, which operate as 50\,$\Omega$ inputs for the RG58 coaxial cables capacitively coupled to the anode of the KamLAND PMTs. 
The analog signal from each PMT is fed into a daughter card, where the signal is terminated and split into \textcolor{black}{$h$-gain and $l$-gain channels}. The $h$-gain output from the daughter cards utilize the RF-ADCs, while the $l$-gain output employs additional 250\,MS/s, 16-bit ADCs (ADS42LB69). The use of the slow DACs enables the RF-ADCs to digitize -150\,mV to +50\,mV in the input level with a $\sim 50$~$\mu$V resolution. \textcolor{black}{The $l$-gain resolution is 0.13\,mV and its dynamic range is set to an input equivalent voltage of -8\,V to +0.5\,V. A few tens of mV pulses can be used to check the consistency between both gain channels.} To further minimize reflections off of the inputs to the amplifier, we also back terminate the input at the PMT base as well. 
The front panel also includes two LEMO connectors, designated for Transistor-Transistor Logic (TTL) input and output. These TTL interfaces are used for external-triggered waveform recording and debugging.  

The rear face of the board houses two VME connectors, two RJ45 connectors, and two enhanced small form-factor pluggable (SFP+) ports. 
The VME connectors are used for providing power ($\pm 12$\,V, 5\,V, 3.3\,V). 
One RJ45 port manages slow control and data transmission via 1~GbE, 
while the other functions as an input for an external clock (50\,MHz) and system reset for synchronization with external modules. 
The SFP+ ports are provisioned for prospective expansions and can potentially facilitate 10\,GbE data transmission. 



\subsection{Analog daughter card}\label{subsec:analog}
The analog circuit on the daughter card amplifies the uni-polar PMT signal to detect single-to-a-few-tens \textcolor{black}{PE}  
through the $h$-gain channel and 
attenuates the signal to record large pulse waveforms 
induced by the cosmic-muon events through the $l$-gain channel. Both channels are designed to have a gain response as flat as possible. Noise levels are minimized through the choice of low-noise amplifiers, anti-aliasing via the natural frequency response of the amplifiers, electromagnetic induction shielding, and careful attention to circuit board layout. Additionally, the digitized signal on the RFSoC PL further reduce noise through oversampling.

\begin{figure}
\centering
\includegraphics[width=1\columnwidth]{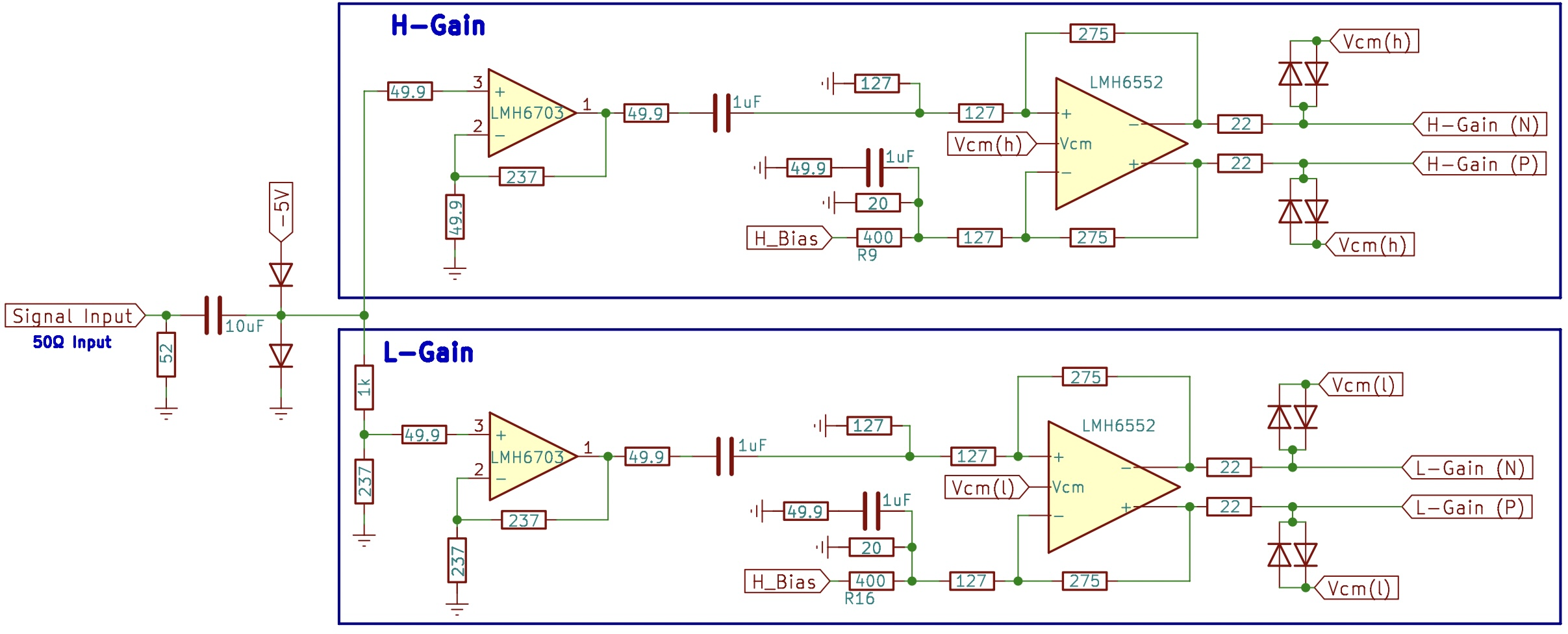}
\caption{A simplified schematic of the analog daughter card design described in the text. 
}
\label{fig:analog}
\end{figure}

The $h$-gain signal path passes through an isolation resistor to a cascade of non-inverting LMH6703 (SOT23-6 package) ultra-low distortion wide-bandwidth current feedback amplifiers. 
The output of the amplifiers is 
capacitively coupled to a single ended input fully differential amplifier (FDA). A current feedback LMH6552 (WSON package) FDA is used to drive the inputs of the RF-ADC. The FDA 1.2\,V common mode voltage is supplied by the RF-ADC. Given the uni-polar nature of the PMT signal, we unbalance the feedback path of the FDA such that we can maximize our dynamic range. 
When the input to the front end is grounded, the common mode voltage of the FDA is 1.2\,V and the unbalancing sets the default output levels to approximately 1.4\,V and 1.0\,V. This maximizes the dynamic range and provides 50\,mV input equivalent of overhead for digitizing the overshoot introduced through capacitive coupling. A clamping diode on each output of the FDA limits the voltage between 1.2\,V +/- 0.4\,V. This is well within the fully reliable region (-0.3\,V to 2.1\,V) of the RF-ADC and within the RF-ADC over-voltage range (1.2\,V +/-0.45V).  The conceptual analog design is in Fig.~\ref{fig:analog}. 

The $l$-gain channel is designed to digitize the $\sim$8\,V pulse with a 0.13~mV resolution. 
The analog circuit attenuates the signal to effectively match the dynamic range of the $l$-gain ADCs. 
To check the consistency between the $h$-gain and the $l$-gain, 20-70\,\textcolor{black}{PE} pulses should be effectively recorded with both gain channels. 
The $l$-gain channel is standard analog design with conventional ADCs. In this paper, we do not report the details of the $l$-gain channel. 

Prior to commissioning the daughter cards for experimental use, the cards are tested by injecting approximately 100~years of equivalent large amplitude pulses into the system and the performance of the cards is re-checked.


\subsection{Programmable logic}
\begin{figure}
\centering
\includegraphics[width=160mm]{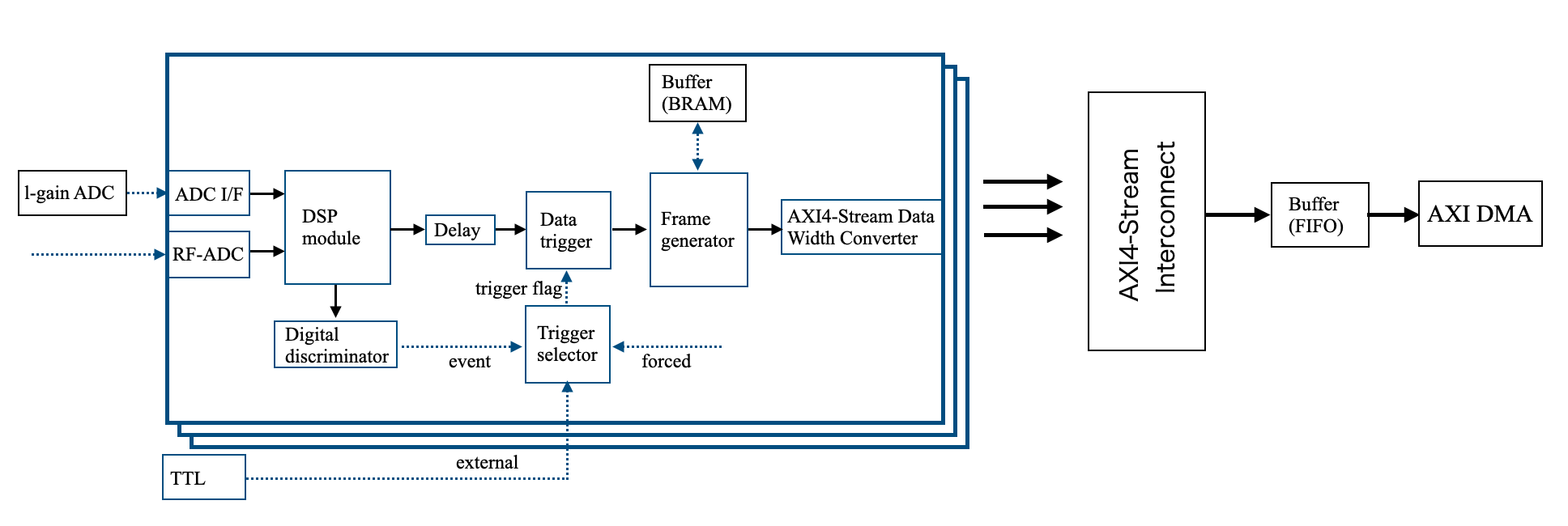}
\caption{PL firmware design}\label{fig:firmwae}
\end{figure}

The data flow through the PL is shown in Figure~\ref{fig:firmwae}. 
We configure the RF-ADCs to operate at 2\,GS/s with a $\times 2$ decimation, effectively yielding an overall sampling rate of 1\,GS/s. 
Each RF-ADC output adheres to the Advanced eXtensible Interface 4 stream~(AXI4-stream) standard. 
Its data size (TDATA) amounts to 128~bits (8~samples $\times$ 16 bits) at a 125\,MHz frequency. 
Its valid signal (TVALID) is concurrently generated by the RF-ADCs. 
Notably, Xilinx guarantees the upper 12 bits of the RF-ADCs' output.  
The lower 4~bits are truncated, and the remaining bits are routed through the digital signal processing (DSP) module. 
This module applies filtering to enhance the signal-to-noise ratio for single photoelectron measurements and correct overshoot phenomena following large pulse events. 

The DSP module's output is split into two AXI4-Streams. 
One stream is directed to a digital discriminator, which issues an event-hit flag based on a simple amplitude threshold. 
Alternative types of hit flags include an external-hit flag derived from the TTL input and a forced flag controlled by the PS and/or an external module via the RJ45 connector. 
These flags converge as a single trigger flag at a trigger selector, which amalgamates them according to the designated trigger mode.

The other stream is delayed to accommodate the processing time of the digital discriminator and trigger selector. 
On the data trigger, the TVALID signal within this AXI4-Stream is then temporarily set to 'Low' ('L'). 
Determined by the trigger flag outcome from the trigger selector, TVALID is subsequently set to 'High' ('H'), inclusive of pre-defined post/pre times. 
A frame generator identifies individual events based on the state of TVALID and produces a corresponding frame of data inclusive of headers, footers, timestamps, and $l$-gain ADC values 
using the 56\,kB buffer located on block random access memories (BRAM). 
The output from the frame generator is converted to an AXI4-Stream.  

The 16 AXI4-Streams from the 16 frame generators are managed via an AXI4-Stream Interconnect, a feature provided by Xilinx employing the Round-Robin arbitration scheme. 
The merged AXI4-Stream is subsequently directed towards the AXI Direct Memory Access (AXI DMA) Intellectual Property (IP) module via a First in First out~(FIFO) buffer. 
All pertinent parameters governing PL functions are set through AXI4-lite or AXI4 on the PS. 

\subsection{DSP module}
\begin{figure}
\centering
\includegraphics[width=140mm]{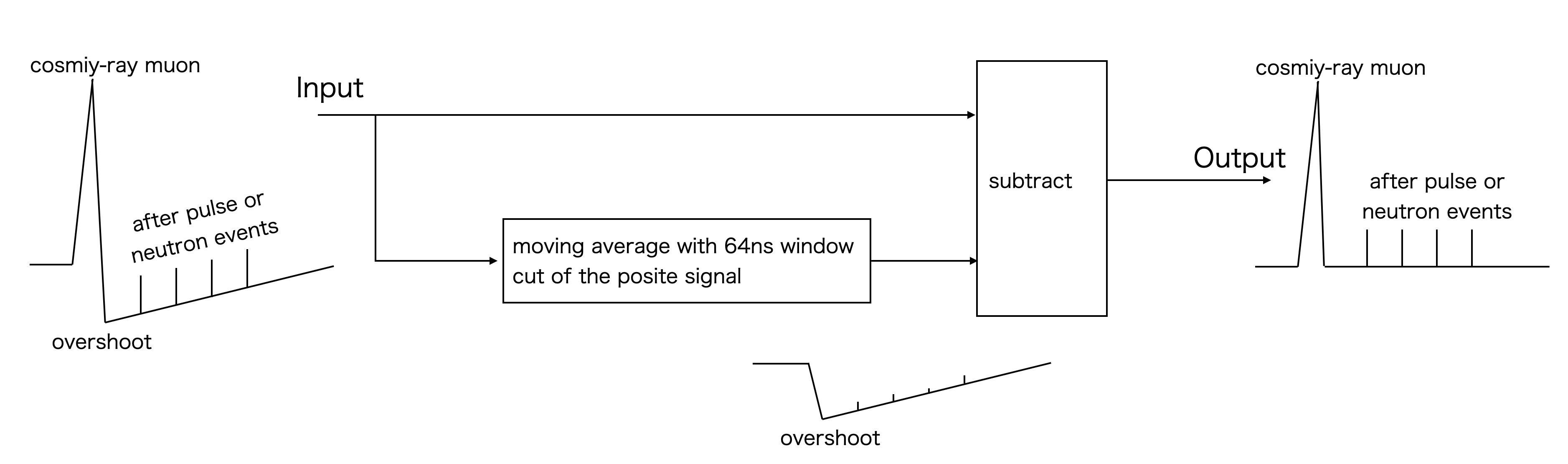}
\caption{Concept of digital BLR}\label{fig:blr_concept}
\end{figure}

A DSP module performs baseline restoration~(BLR) and digital filtering. 
The BLR aims to remove the waveform overshoot after a large pulse. This overshoot is effectively mitigated by computing a 64~ns moving average within the negative region of the waveform. Subtracting this moving average from the original digitized waveform results in the generation of the BLR waveform. A conceptual image of BLR is shown in Fig.~\ref{fig:blr_concept}. 

The BLR waveform is now digitally filtered by applying a simple 8\,ns moving average~(MA). 
The selection of an 8\,ns length is optimized to minimize high-frequency noise while preserving the PMT signal. This design is developed using System Generator for DSP and MATLAB/Simulink.

\subsection{Processing system}
The PS, encompassing a quad-core Arm Cortex-A53 architecture, operates on a Linux environment that has been compiled using Petalinux 2019.2. The PS reads the data from PL and sends them to back-end computer(s) through a 1~GbE connection, which is implemented via the Gigabit Ethernet MAC (GEM3) module employing Multiplexed Input/Output (MIO). To facilitate seamless communication via traditional TCP sockets, the system has been equipped with ZeroMQ~\cite{zeromq}. Consequently, the main board can function as a ZMQ module, simplifying communication tasks. It's worth noting that a potential future enhancement includes the possibility of utilizing the 10~GbE connection with high-speed transceivers in the PL.



\section{Prototype test}
We conducted an evaluation of our customized electronics using a Wiener VME9U crate. 
The electronics demonstrated successful operation with the following power specifications: +12\,V at 0.7\,A, -12\,V at 0.3\,A, +5\,V at 6\,A, and +3.3\,V at 0.4\,A. 
This translates to a total power consumption of 43\,W with the test firmware in place. 
We note that the power consumption of the +3.3\,V line is expected to increase when utilizing the 10~GbE connection, since it is primarily utilized to power SFP+ modules.


In comparison, our previous dead-time free electronics, which utilized the conventional ADCs and FPGA configuration, operated with 12 analog inputs and consumed 74\,W. This represents an enhanced power efficiency per channel of 56\%. 
In this section, we delve into board testing, with a primary focus on the $h$-gain channels.

\subsection{Frequency response}
\begin{figure}[htbp]
\centering
\includegraphics[width=100mm]{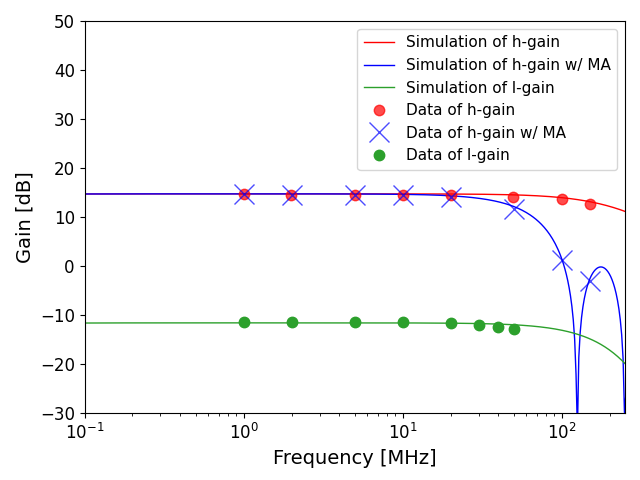}
\caption{Frequency response of $h$-gain with and without the 8~ns moving average~(MA) on DSP and $l$-gain. There is no significant difference between data and simulation. }
\label{fig:freq_response} 
\end{figure}

We conducted measurements to evaluate the response using 10\,mV (500\,mV) sine waves across various frequencies within the $h$-gain ($l$-gain) channels. 
Figure~\ref{fig:freq_response} provides a comparison between the measured frequency responses and the expected responses, with MA on the DSP module enabled (disabled).
With MA, the designed cut-off frequency is set at 68~MHz. The measured and designed frequency responses exhibit no significant disparities.
This result shows that the RF-ADC's frequency response maintains flatness within the range of 1 to 125~MHz, with a cut-off frequency controllable through the DSP module. 
\textcolor{black}{We confirmed that the total frequency response is wider than the KamLAND2 requirements~(60\,MHz).}

\subsection{Noise level}
\begin{figure}[htbp]
\centering
\includegraphics[width=140mm]{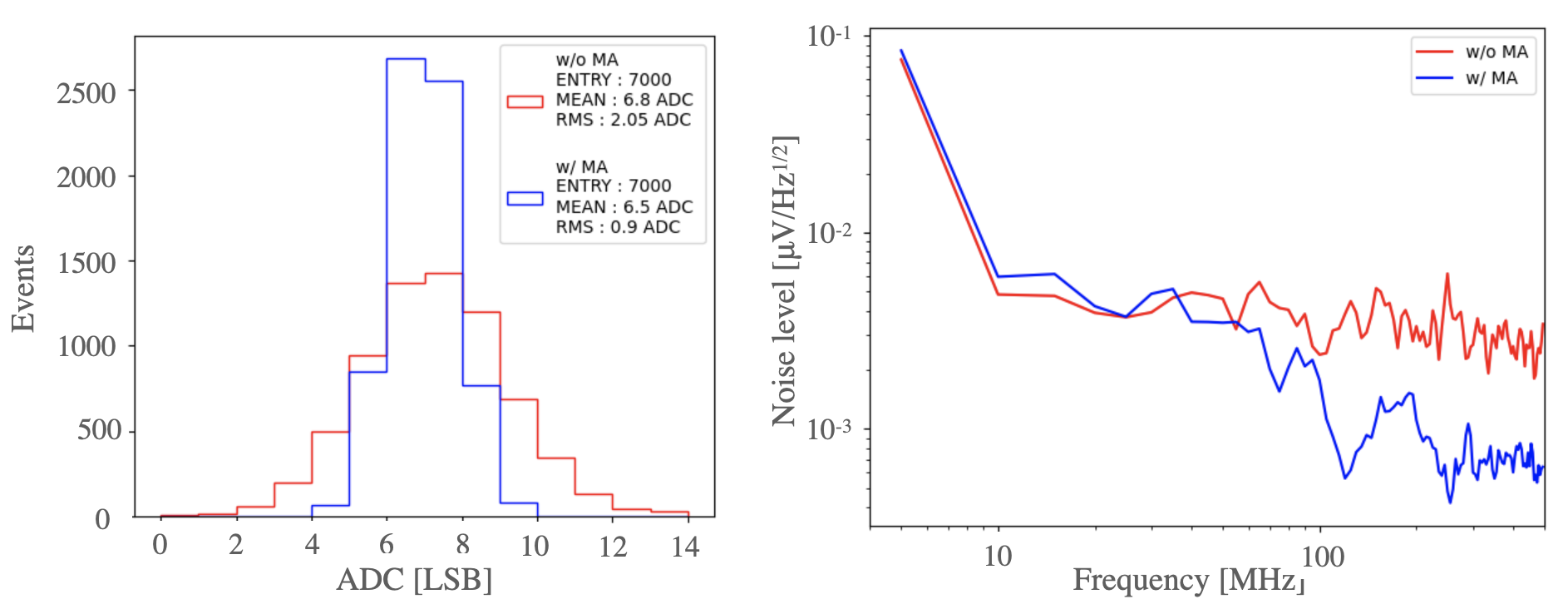}
\caption{Noise histogram~(left) and spectrum~(right) with and without \textcolor{black}{MA}. Applying \textcolor{black}{MA}, the noise level is 0.9\,ADC$_\textrm{rms}$ and there is no structure in the frequency space.}
\label{fig:noise_spec} 
\end{figure}

\begin{figure}
\centering
\includegraphics[width=100mm]{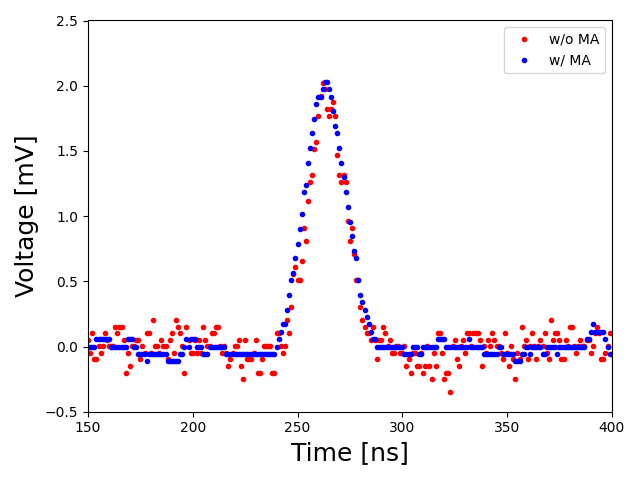}
\caption{Example of 2\,mV pulse detection with and without 8\,ns moving average~(MA). An amplitude of 2~mV corresponds to the SPE peak. }\label{fig:SPE_pulse}
\end{figure}

We performed measurements to assess the noise level in the absence of input configuration. 
Figure~\ref{fig:noise_spec} shows both the histogram and spectrum of the input equivalent noise from the $h$-gain data. 
Without MA, the root mean square (RMS) noise level amounts to 2.1~ADC$_\textrm{rms}$, which corresponds to an input equivalent 100~$\mu$V$_\textrm{rms}$. 
A noise spectral density measurement indicated the line noise at 250\,MHz, which corresponded to double the frequency of the system clock. 
Employing MA on DSP suppress the noise level to 0.9 ADC$_\textrm{rms}$~(an input equivalent 40~$\mu$V$_\textrm{rms}$). 
The right panel of Fig.~\ref{fig:noise_spec} depicts how the MA 
effectively mitigates the line noise. 
With this low noise level, 2~mV pulses, which 
corresponds to a single \textcolor{black}{PE} pulse, are clearly visible, as shown in Fig.~\ref{fig:SPE_pulse}.

\subsection{Discriminator threshold}
\begin{figure}[htbp]
\centering
\includegraphics[width=100mm]{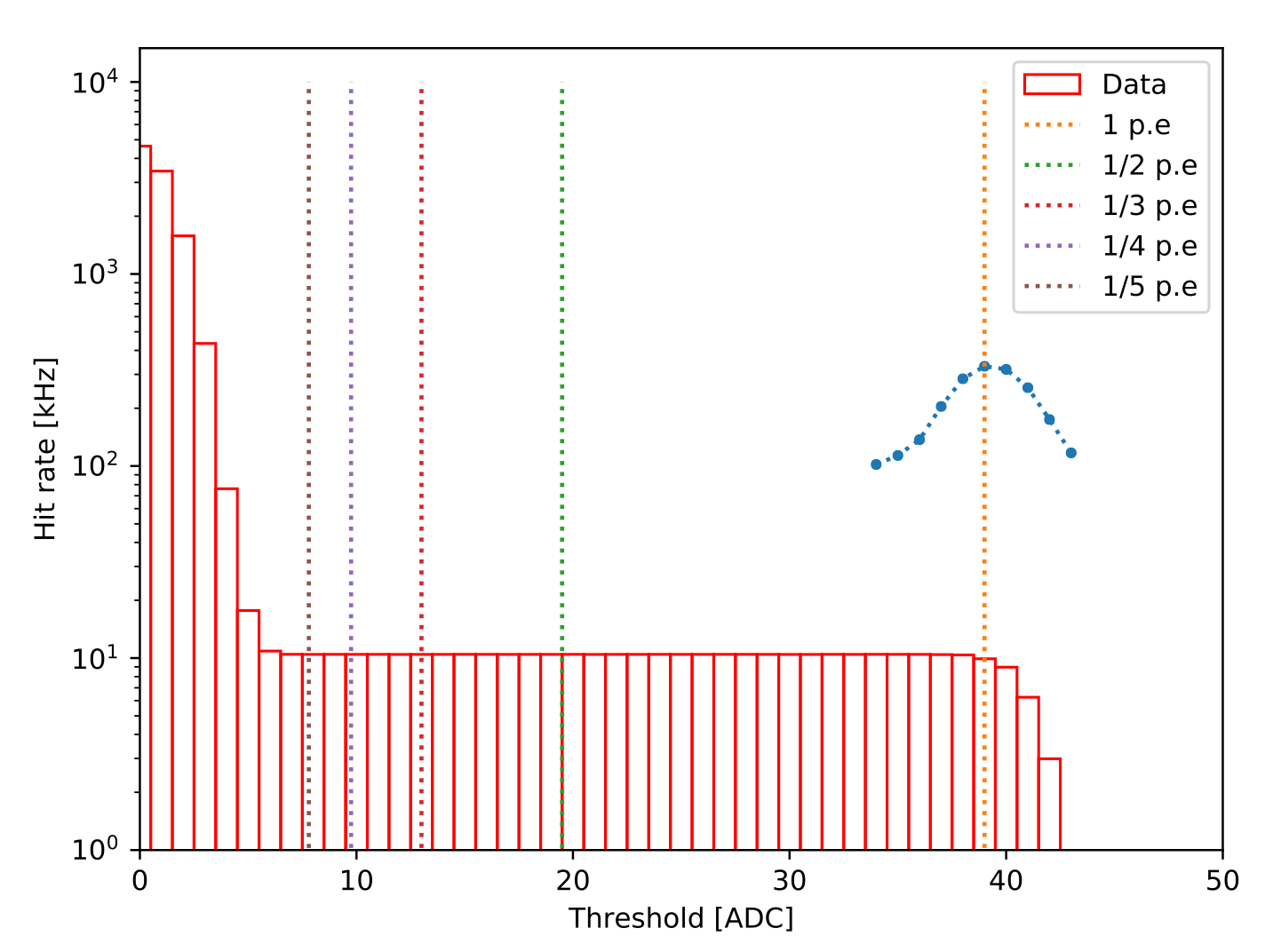}
\caption{Results of threshold scan using 2\,mV pulses with a duration of 20\,ns. From its derivative~(blue dotted curves), a clear peak structure can be seen around threshold of 39\,ADC counts. Given the scan, a threshold can be safely set at the 1/5\,\textcolor{black}{PE}. level. }\label{fig:threshold_scan}
\end{figure}

A threshold scan was conducted in order to evaluate the discriminator thresholds available.
Figure~\ref{fig:threshold_scan} illustrates the hit rate as a function of discriminator thresholds. 
For this evaluation, \textcolor{black}{
2\,mV~triangular pulses, with 
a rising time of 20~ns and a falling time of 20~ns,} were injected into the system using an arbitrary waveform generator at a frequency of 1\,kHz. 
Given that we expect the single \textcolor{black}{PE} peak to occur around 40\,ADC counts, a threshold of 1/5\,\textcolor{black}{PE} is achievable. 
Regrettably, the noise originating from the waveform generator amounts to 126\,$\mu$V$_\textrm{rms}$, nearly three times larger than the electronics noise. \textcolor{black}{Without the noise from the function generator, a 1/6\,\textcolor{black}{PE} threshold might be feasible.}

\subsection{Linearity}
\begin{figure}
\centering
\includegraphics[width=140mm]{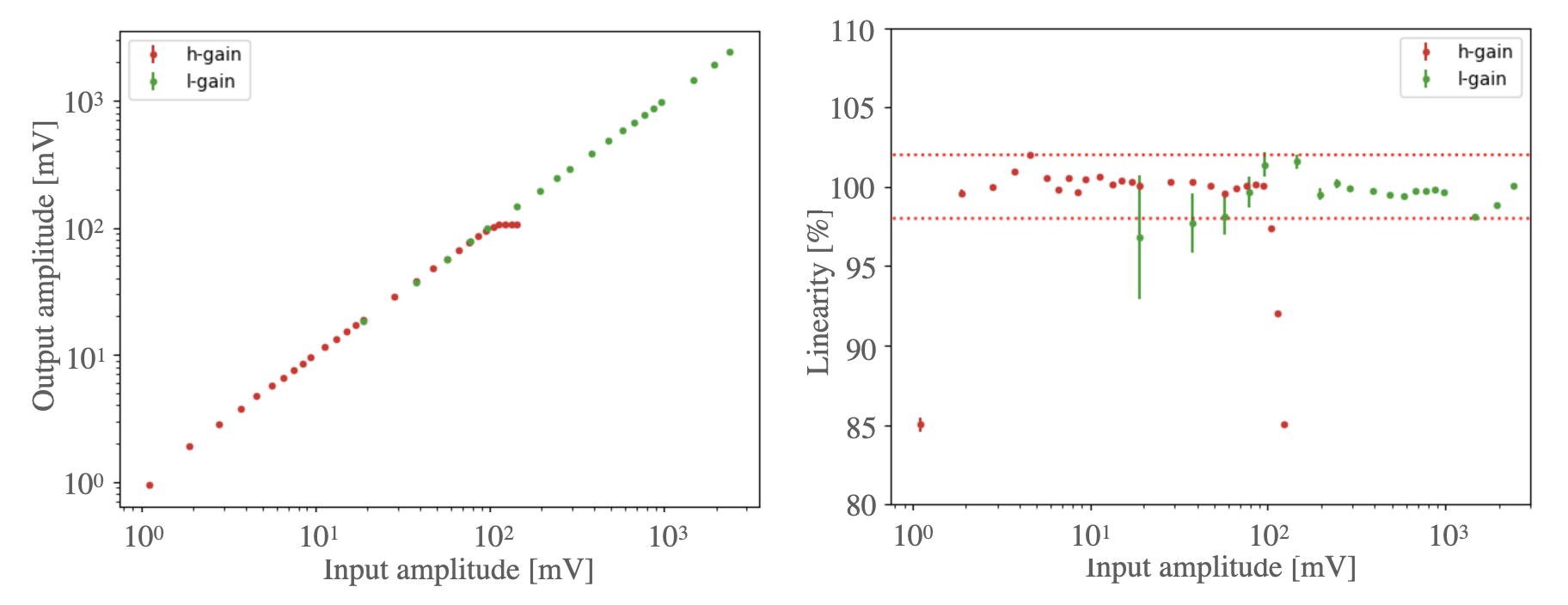}
\caption{(Left) Output (recorded) amplitudes as a function of input amplitudes. (Right) Residual as a function of input amplitude. There is sufficient overlap range between $h$-gain and $l$-gain. 
Input and output amplitudes match within 2\% most of the range.}\label{fig:linearity}
\end{figure}

Ensuring linearity in the response is crucial when quantifying the number of \textcolor{black}{PE} detected by PMTs. 
To assess this aspect, we employed a waveform generator to generate pulses characterized with a 40\,ns~time window. 
Our evaluation involved comparing the input equivalent amplitudes against the corresponding input amplitudes across various levels~(Fig.~\ref{fig:linearity}). 
Barring pulses less than 1\,mV and exceeding 100\,mV, the input equivalent amplitude is observed to be withing $\pm 2\%$ of the desired level. 
Our target is the single \textcolor{black}{PE} events, typical amplitude $\sim 2$\,mV. Thus, deviations at 1\,mV is not a problem and can be compensated for during calibration if needed.
By tuning the DAC output, it is feasible to tune the saturation amplitude to approximately 150\,mV. 
Importantly, the $h$-gain and $l$-gain are naturally interconnected. 

\subsection{Crosstalk}
\begin{figure}
\centering
\includegraphics[width=100mm]{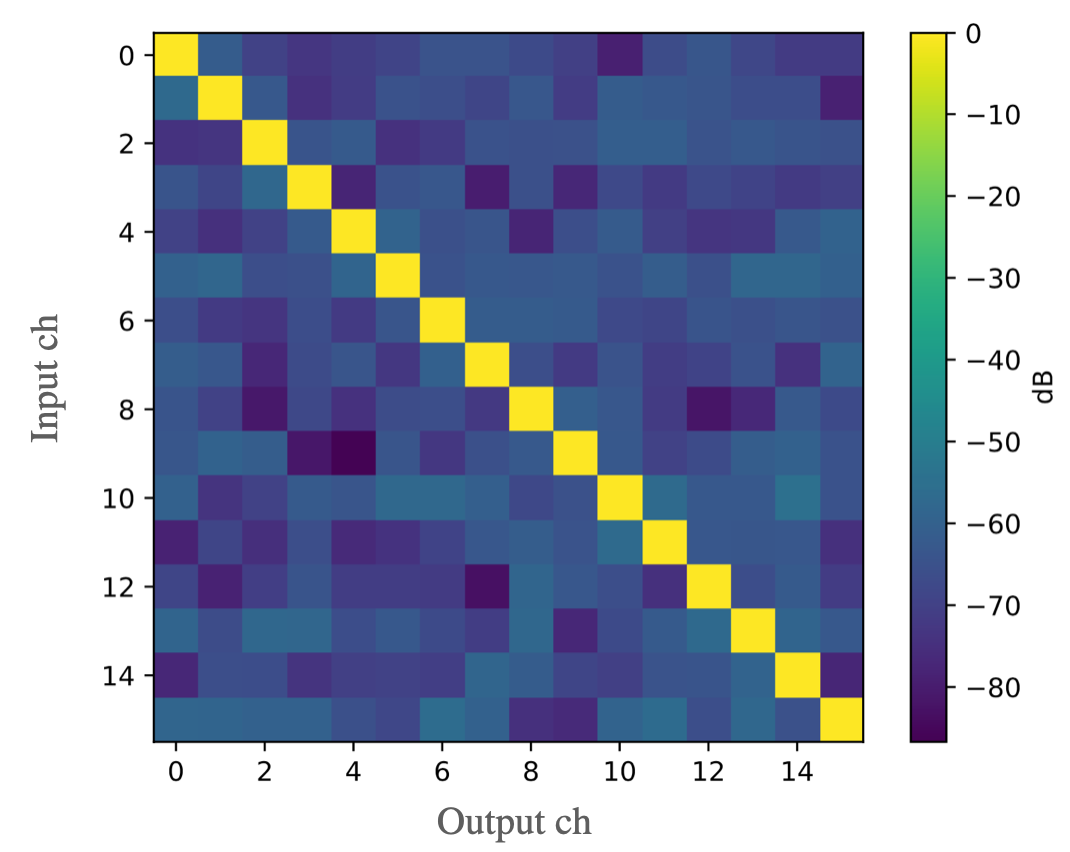}
\caption{Results of crosstalk measurements. 
The vertical axis shows the input channel of the sine wave, the horizontal axis shows the channel of the crosstalk measurement at that time, and the $z$ axis shows the magnitude of the crosstalk.
}\label{fig:cross-talk}
\end{figure}

To prevent the creation of spurious single photoelectron-like signals, a crosstalk level of at \textcolor{black}{most} -55 dB is necessary. 
To evaluate the crosstalk, we introduced sine waves with a frequency of $f = 10$~MHz and an amplitude of 25~mV. 
We define the "rock-in amplitude" ($A$) as: 
\begin{align}
A=2\sqrt{\langle s(t) \times \cos(2\pi f t) \rangle^2 + \langle s(t) \times \sin(2\pi f t)  \rangle^2)} 
\end{align}
$\langle \cdot \cdot \cdot \rangle$ is mean of $\cdot \cdot \cdot$, $s(t)$ is measured time-series data as a func of time $t$.  
When the sine waves is introduced to $j$ channel, evaluating the crosstalk from channel $j$ to channel $i$ involves the calculation:
\begin{align}
    20 \log_{10} \Biggl( \frac{A^i}{A^j} \Biggr)~\textrm{[dB]},
\end{align}
where $i = 0,1,\cdot \cdot \cdot 15$ but not $j$.  
Fig.~\ref{fig:cross-talk} shows the measured crosstalk levels, showcasing that they remain below the required level. 
The average crosstalk was measured at -62.3~dB. 
These measurements indicates that the observed crosstalk is within acceptable limits, effectively preventing the generation of fake single photoelectron signals.

\subsection{Test with a real 20-inch PMT}
\begin{figure}
\centering
\includegraphics[width=120mm]{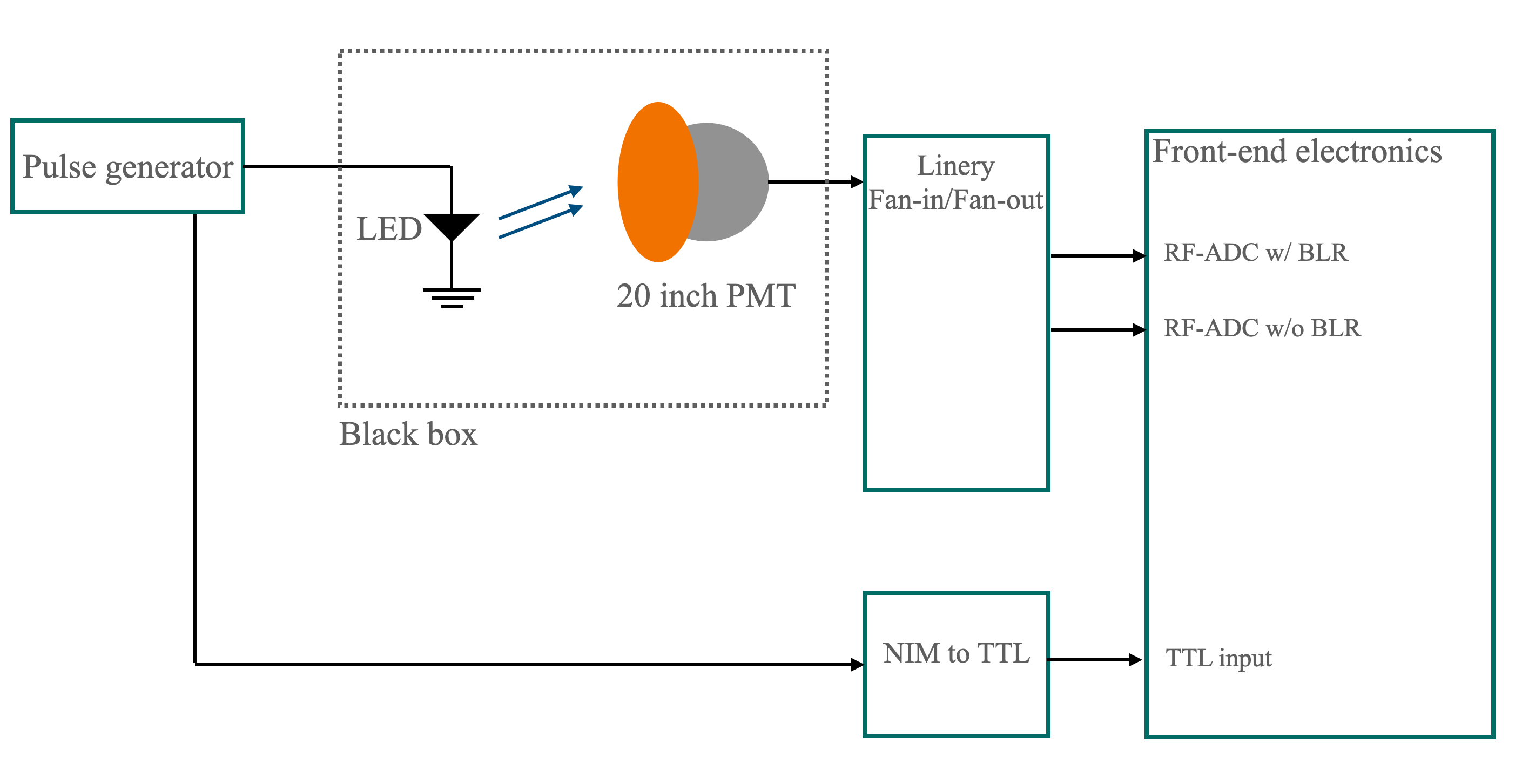}
\caption{Test setup with a read 20-inch PMT. }\label{fig:blr_test}
\end{figure}

We validated the baseline correction function implemented within BLR on the DSP module using a 20-inch PMT~(HAMAMATSU R12860). 
This validation process involved employing a light-emitting diode (LED) to illuminate the PMT, as depicted in Fig.~\ref{fig:blr_test}. 
The timing of LED's light was controlled through a pulse generator. 
The trigger output from the generator signal was supplied to the front-end electronics for the external-hit flag.

\begin{figure}
\centering
\includegraphics[width=100mm]{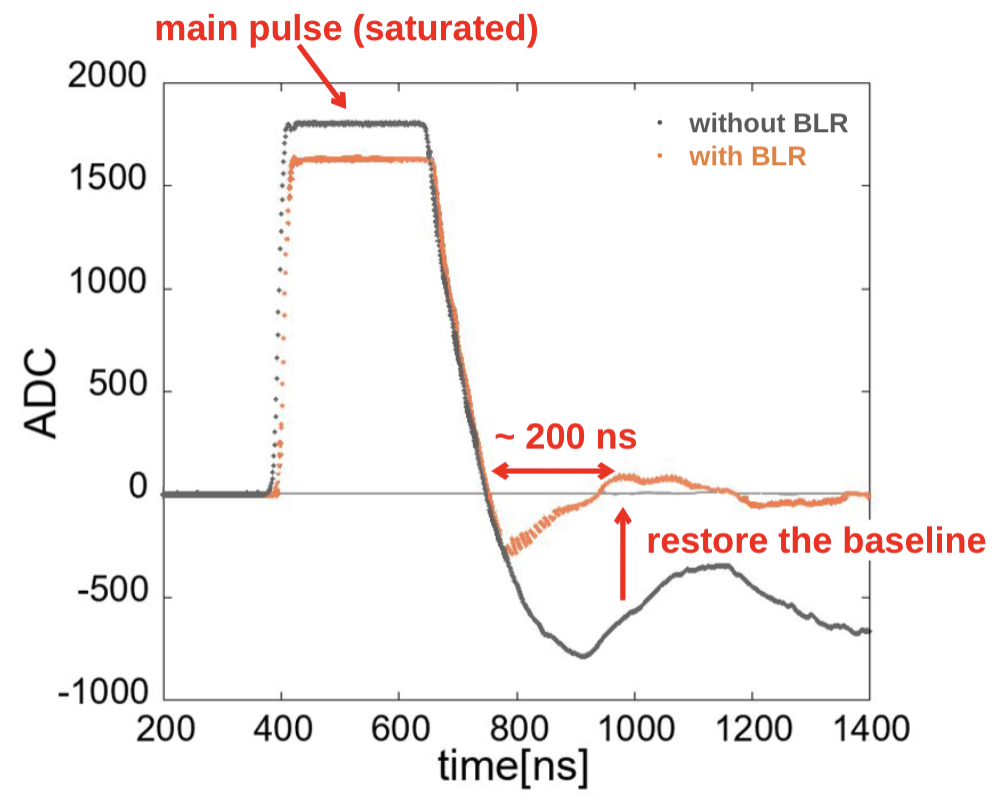}
\caption{LED pulse and overshoot waveforms w/ and w/o the BLR on the DSP module. 
With BLR, the baseline restored 200~ns after the end of the LED pulse. }\label{fig:blr_result}
\end{figure}

Fig.~\ref{fig:blr_result} shows the acquired waveform data both with and without BLR. The baseline was subsequently adjusted to zero during the offline analysis. 
The primary LED-induced pulses appears around the 400-600~ns. 
Without BLR, the baseline deviates from zero due to the presence of overshoot.  
With BLR, the baseline effectively returns to zero within approximately 200~ns.  
We also confirmed that the baseline correction did not cause any distortion for the PMT signal waveform. 

\subsection{Data transfer}

\begin{figure}
\centering
\includegraphics[width=100mm]{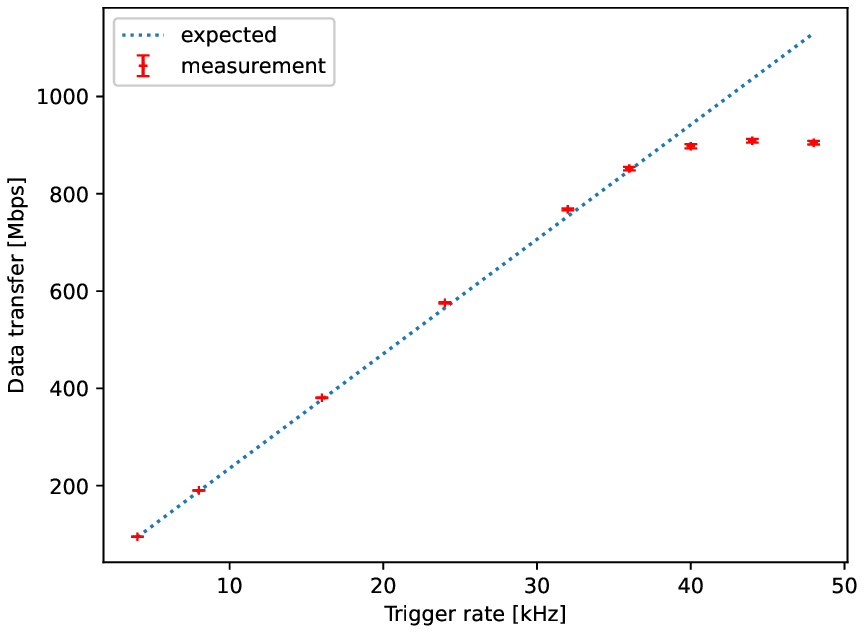}
\caption{Measured and expected data transfer speed by varying the forced trigger rate. 
This evaluation was conducted using a ZMQ client on a back-end computer. 
The measured and expected data transfer speeds aligned well up to 32~kHz.}\label{fig:dataspeed}
\end{figure}

We evaluated the data transfer speed with respect to the external-hit flag rate. 
Our approach involved setting up a ZMQ client on a back-end computer running Ubuntu 18.04.6 LTS. 
This client was designed to receive data from the ZMQ server located on the RFSoC PS. 
Throughout this evaluation, the focus was solely on measuring the data transfer speed within the ZMQ client. 
To maintain the clarity of the assessment, we refrained from implementing additional processes like data saving. 
The results of this evaluation are presented in Fig.~\ref{fig:dataspeed}, 
showcasing the measured data transfer speed versus varying external trigger rates. 
The figure indicates consistency, with expected values and stable data transfer performance for external triggers below 32~kHz~(750\,Mbps). 
However, a noticeable deviation becomes apparent when comparing the measured and expected speeds beyond an external trigger rate of 40\,kHz. 
These findings allow us to validate that a dark rate of 25~kHz and a data transfer speed of 600\,Mbps are attainable for the intended purposes in KamLAND. 
It should be noted that throughout the entirety of this evaluation, our utilization was restricted to the capabilities of 1\,GbE. 
The incorporation of 10\,GbE through the SFP+ connector will open the potential for conducting data-taking operations at significantly higher event rates. 


\section{Summary}
RFSoC technology represents a cutting-edge advancement in signal processing and data acquisition. This paper has been dedicated to the development of RFSoC-based front-end electronics for pulse detection, accompanied by an extensive feasibility study.
The prototype has demonstrated well-understood frequency response, low noise~($\sim 40$\,$\mu$V$_\textrm{rms}$), linearity within $\pm$2\%, low crosstalk~(<55\,dB), and stable data transfer up to the 36~kHz trigger rate~(750\,Mbps) \textcolor{black}{with 16 inputs using 43\,W.}
The DSP module has proven effective in correcting the PMT overshoot. 
Moreover, the system's power consumption has demonstrated a notable advantage over the traditional FPGA + ADC system. 
The results of this study provide robust evidence in favor of the feasibility and potential of adopting the RFSoC-based approach for pulse detection. With some necessary modifications, the front-end electronics described here hold promise for practical implementation in the KamLAND\textcolor{black}{2} system.





\acknowledgments
This work was supported by JSPS KAKENHI Grant Number 19H058028, the National Science Foundation (NSF) No.~NSF-1806440,~NSF-2012964, the Heising-Simons Foundation, as well as other DOE and NSF grants to individual institutions, U.S.-Japan Science and Technology Cooperation Program in High Energy Physics. 
We would like to say thank J.~Suzuki for useful discussion about RFSoC. 
The authors are also grateful to all KamLAND collaborators for supporting this effort.


\bibliographystyle{JHEP}
\bibliography{rfsoc_mogura}

\end{document}